\documentclass[%
 aip,
 amsmath,amssymb,
 reprint,%
]{revtex4-1}
\usepackage{color}
\usepackage{graphicx}
\usepackage{dcolumn}
\usepackage{bm}
\usepackage{CJK}

\allowdisplaybreaks

\begin{document}


\preprint{AIP/123-QED}


\title{Anomalous behavior of trapping in extended dendrimers with a perfect trap}

\author{Zhongzhi Zhang}
\email{zhangzz@fudan.edu.cn}
\homepage{http://www.researcherid.com/rid/G-5522-2011}

\author{Huan Li}

\author{Yuhao Yi}

\affiliation {School of Computer Science, Fudan University,
Shanghai 200433, China}

\affiliation {Shanghai Key Laboratory of Intelligent Information
Processing, Fudan University, Shanghai 200433, China}

\date{\today}

\begin{abstract}
Compact and extended dendrimers are two important classes of dendritic polymers. The impact of the underlying structure of compact dendrimers on dynamical processes has been much studied, yet the relation between the dynamical and structural properties of extended dendrimers remains not well understood. In this paper, we study the trapping problem in extended dendrimers with generation-dependent segment lengths, which is different from that of compact dendrimers where the length of the linear segments is fixed. We first consider a particular case that the deep trap is located at the central node, and derive an exact formula for the average trapping time (ATT) defined as the average of the source-to-trap mean first passage time over all starting points. Then, using the obtained result we deduce a closed-form expression for the ATT to an arbitrary trap node, based on which we further obtain an explicit solution to the ATT corresponding to the trapping issue with the trap uniformly distributed in the polymer systems. We show that the trap location has a substantial influence on the trapping efficiency measured by the ATT, which increases with the shortest distance from the trap to the central node, a phenomenon similar to that for compact dendrimers. In contrast to this resemblance, the leading terms of ATTs for the three trapping problems differ drastically between extended and compact dendrimers, with the trapping processes in the extended dendrimers being less efficient than in compact dendrimers.
\end{abstract}

\pacs{36.20.-r, 05.40.Fb, 05.60.Cd}

\maketitle



\section{Introduction}

As an integral class of macromolecules~\cite{GuBl05}, dendrimers have attracted extensive attention of scientists in the interdisciplinary fields of physics, chemistry, and materials, since they display unusual geometrical, physical, and chemical properties~\cite{ToNaGo90,Fr94,JiAi97,KoShShTaXuMoBaKl97}. These chemical compounds are synthesized by repeating units arranged in a hierarchical self-similar fashion, and are characterized by basic building element, branching of the end groups, as well as the number of generations. Among various dendritic polymers, compact and extended dendrimers constitute two important types of dendrimer families. In the compact family, the length of linear unit (segment) in all generations is identical, while in the extended family segment length is varying, which decreases toward the periphery. The unique structure of two dendrimers makes them potentially promising candidates for a broad ranges of applications, e.g., artificial antenna systems for light harvesting~\cite{KoShShTaXuMoBaKl97,BaCaDeAlSeVe95}.

Light harvesting by dendrimers can be described by trapping process~\cite{BaKlKo97,BaKl98,BaKl98JPC,BaKl98JOL,PeZh14}, which is a kind of random walks with a deep trap positioned at a given site, absorbing all particles visiting it. The highly desirable quantity for this paradigmatic dynamical process is trapping time (TT), also known as mean first-passage time (MFPT)~\cite{Lo96,Re01,NoRi04,CoBeMo05,CoBeKl07,CoBeMo07,CoBeTeVoKl07}, which is the expected time for a particle starting off from a source point to first reach the trap. The average trapping time (ATT), defined as the average of trapping time over all starting nodes, gives insight to the trapping process, since it provides a quantitative measure of trapping efficiency. In addition to light harvesting, trapping is also closely related to many other significant dynamical processes in diverse complex systems, such as target search~\cite{JaBl01,BeCoMoSuVo05,Sh06,AgBlMu10,BeLoMoVo11,HwLeKa12,HwLeKa12E}, energy or exciton transport in polymer systems~\cite{SoMaBl97,BlZu81,MuBlAmGiReWe07,AgBlMu10IJBC,Ag11,MuBl11}.

In consideration of its direct relevance, trapping in various complex systems has been an active subject of research in the past years, with particular attention focused on determining ATT in diverse systems that display different structural and other properties. Thus far, trapping problem has been extensively studied for many systems, including regular planar and cubic lattices~\cite{Mo69,GLKo05,GLLiYoEvKo06}, $T-$ fractals and their extensions~\cite{KaRe89,Ag08,HaRo08,LiWuZh10,ZhWuCh11,JuWuZh13,PeXu14}, Sierpinski gasket~\cite{KaBa02PRE,BeTuKo10} and Sierpinski tower~\cite{KaBa02IJBC}, small-world uniform recursive trees~\cite{ZhQiZhGoGu10,ZhLiLiCh11,LiuZh13}, scale-free networks~\cite{KiCaHaAr08,ZhQiZhXiGu09,ZhXiZhLiGu09,TeBeVo09,AgBu09,AgBuMa10,ZhLiMa11,MeAgBeVo12,YaZh13}, complex weighted networks~\cite{ZhShCh13,LiZh13,LiZh14}, as well as directed weighted treelike fractals~\cite{WuZh13}. These studies uncovered how the behavior of ATT for trapping is affected by different properties of complex systems, such as topological structure, weight and direction of edges.

Except for the aforementioned systems, concerted theoretical efforts have also been devoted to trapping problem in polymer systems (especially dendrimers), reporting specific properties and phenomena of the trapping process in these macromolecules. The trapping problem in compact dendrimers was first addressed in Refs.~\cite{BaKlKo97,BaKl98,BaKl98JPC,BaKl98JOL,PeZh14}, where the MFPT from the peripheral node to the central node was computed both in the presence and in the absence of a fixed energy bias. Partly inspired  by these works, applying the theory of finite Markov chain, the exact expression for MFPT from an arbitrary node to the central node was derived in Refs.~\cite{BeHoKo03,BeKo06,WuLiZhCh12}, where the ATT was also obtained when an immobile trap is located at the center. The influence of trap location on ATT for trapping in compact dendrimers was explored in~\cite{LiZh13JCP}. Moreover, the factors governing the trapping efficiency for compact dendrimers have also been much studied in detail~\cite{RaShChMu00,RaShChMu02,RaGa03,HeMaKn04,FlAmShKl06}.

In contrast to compact dendrimers, trapping problem in extended dendrimers has received less attention~\cite{BeHoKo03}, although they are an important class of nanoscale supermolecules and their specific structure is suggested to yield different trapping behavior from that for their compact counterparts. Since extended dendrimers are an important class
of nanoscale supermolecules, it is of theoretical and practical
importance to address the influences of topology of
extended dendrimers on trapping occurring on them.

In this paper, we present a comprehensive study of trapping in extended dendrimers, with an aim to explore the impact of their geometrical structure on the trapping efficiency. We first focus on a special case of trapping problem with a single trap located at the central node. We derive closed-form expressions for the MFPT from an arbitrary node to the trap, as well as for the ATT to the trap over all starting points. Then, using these results, we determine an exact solution for the ATT to an arbitrary target. Finally, we attack the case of trapping with the trap distributed uniformly over all nodes. By using two different techniques, we deduce an explicit formula for the ATT. We show that the trap position has a strong effect on the efficiency of trapping occurring on extended dendrimers, since their leading scalings display distinct dependence on the system size, which increase with the shortest distance from the trap to the central node. We also demonstrate that for all the cases of trapping problems, the dominant behaviors for trapping in extended dendrimers are different from those corresponding to compact dendrimers.

\section{Construction and properties}

In this section, we introduce the construction of extended dendrimers and study their relevant properties.

\subsection{Construction method}

\begin{figure}
\begin{center}
\includegraphics[width=0.8\linewidth,trim=0 0 0 0]{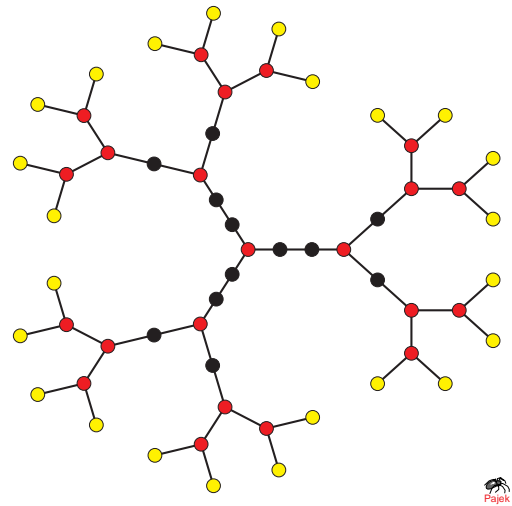}
\caption{Structure of an extended dendrimer network $D_4$. }  \label{Cayley}
\end{center}
\end{figure}

We concentrate on a particular network for extended dendrimers, with the number of nearest neighbors of any branching site being 3. Let $D_g$ $(g \geq 1)$ denote the extended dendrimer having generation number $g$, with the length of segment at generation $i$ being $g-i$. Then, $D_g$ can be built by $g$ iterations in the following way~\cite{RaShChMu00,MaRoSi02}. At iteration $0$, the dendrimer comprises only one central node; at iteration $1$, 3 segments with length $g-1$ are created being attached to the central node. At iteration $i$ ($1<i<g$), for each branching node of generation $i-1$ (i.e., the node generated at iteration $i-1$ farthermost from the center), $2$ segments (chains) containing $g-i$ nodes are generated and are linked to the branching node. At the last iteration $g$, for each node created at generation $g-1$, we add a pair of new nodes and link them to it. All the nodes introduced at this stage are called peripheral nodes of $D_g$. Figure~\ref{Cayley} illustrates the structure for a specific extended dendrimer $D_4$.

Let $M_g$ denote the length of the shortest path from an arbitrary peripheral node to the center. Then,
\begin{eqnarray}
M_g =\sum\limits_{i=1}^{g-1} \,i + \ 1= \frac{1}{2}(g^2-g+2).
\end{eqnarray}
According to their shortest distances to the central node, all nodes in $D_g$ can be divided into $M_g+1$ levels, with the central
node being at level 0, and the peripheral nodes being at level $M_g$.

Let $G_i(g)$ denote the generation (iteration) at which all nodes at level $i$ were created. In order to determine $G_i(g)$, we introduce two quantities $L_i(g)$ and $S_i$, where $L_i(g)$ represents the possible maximal value of levels that nodes created at generation $i$ belong to, and $S_i$ is defined to be
\begin{eqnarray}
S_i &=&
\begin{cases}
1,                           & i=0,  \\
1 + \sum\limits_{j=1}^{i} j, & i \geq 1,
\end{cases} \notag \\ \notag \\
&=& \frac{i^2+i+2}{2}.
\end{eqnarray}
Then, we have
\begin{equation}
L_i(g)=
\begin{cases}
M_g-S_{g-i-1}, & 0 \leq i < g, \\
M_g, & i=g,
\end{cases}
\end{equation}
and
\begin{eqnarray}
G_i(g) =
\begin{cases}
0, & i=0, \\
j \ | \ j\in(M_g-S_{g-j},M_g-S_{g-j-1}], & 0<i<M_g, \\
g, & i=M_g. \\
\end{cases}\notag
\\
\end{eqnarray}

By construction, the number of nodes at level $i$ is
\begin{equation}
N_i(g)=
\begin{cases}
1,                    & i=0, \\
3 \cdot 2^{G_i(g)-1}, & 1 \leq i \leq M_g. \\
\end{cases}
\end{equation}
Thus, the total number of nodes in $D_g$ is
\begin{equation}\label{ng}
N_g = \sum\limits_{i=0}^{M_g} N_i(g) = 9 \cdot 2^{g-1}-3g-2\,.
\end{equation}
And the total number of edges in $D_g$ is
\begin{equation}
E_g = N_g-1 = 9 \cdot 2^{g-1}-3g-3\,.
\end{equation}

\subsection{Structural properties}

We proceed to present some important structural properties of $D_g$, focusing on degree distribution and average path length.

\subsubsection{Degree distribution}

It is obvious that, for nodes in $D_g$, there are only three possible degree values (1, 2, and 3). The degree of any node at levels $L_0(g)$, $L_1(g)$, $\ldots$, $L_{g-1}(g)$ is $3$; the degree of all peripheral nodes (i.e., nodes at level $M_g$) is $1$; and the degree of all other nodes is $2$. Let $\Delta_i(g)$ represent the  number of nodes with degree $i$. Then, it is easy to obtain
\begin{eqnarray}
\Delta_1(g) = N_{M_g}(g) = 3 \cdot 2^{g-1}\,,
\end{eqnarray}
\begin{eqnarray}
\Delta_2(g) &=& \sum\limits_{i=1}^{g-2} \sum\limits_{j=L_{i-1}(g)+1}^{L_i(g)-1} N_j(g) \notag\\
               &=& 3 \cdot2^g-3 \cdot2^{g-1}-3g+1\,,
\end{eqnarray}
and
\begin{eqnarray}\label{Np}
\Delta_3(g) = \sum\limits_{i=1}^{g-1} N_{L_i(g)} = 3 \cdot2^{g-1}-3\,,
\end{eqnarray}
respectively.

\subsubsection{Average path length}

The average path length denotes the average of length for the shortest path between two nodes over all node pairs. Assume that each edge in $D_g$ has a unit length. Then the length of the shortest path from node $i$ to $j$ in $D_g$, denoted by $d_{ij}(g)$, is the minimum length of the path connecting the two nodes. Let $\bar{d}_g$ represent the average path length of $D_g$, defined by
\begin{equation}\label{bard}
\bar{d}_g =\frac{P_{\rm tot}(g)}{N_g(N_g-1)/2}\,,
\end{equation}
where $P_{\rm tot}(g)$ is the sum of $d_{ij}(g)$
over all pairs of nodes, i.e.,
\begin{equation}\label{dsdf}
P_{\rm tot}(g) = \sum_{i< j} d_{ij}(g)\,.
\end{equation}

We continue by showing the procedure of determining $P_{\rm tot}(g)$, which just equals the number of edges in the shortest paths between all pairs of nodes in $D_g$. Instead of counting the edges in the paths, here we count the paths passing through a given edge, and then sum the results of all edges in $D_g$. Let $(i,j)$ be an edge in $D_g$ linking two nodes $i$ and $j$, and let $E_{ij}(g)$ be the number of the shortest paths of different node pairs, which pass through $(i,j)$. Let $N_{i<j}(g)$ denote the number of nodes in $D_g$ lying closer to node $i$ than to node $j$, including $i$ itself. Then, total shortest-path distance can be computed through~\cite{Wi47,DoenGu01}
\begin{eqnarray}\label{dscalc}
P_{\rm tot}(g)&=&\sum_{(i,j)\in D_g}E_{ij}(g)=\sum_{(i,j)\in D_g} N_{i<j}(g)N_{j<i}(g)\nonumber\\
&=& \sum_{(i,j)\in D_g}N_{i<j}(g)[N_g-N_{i<j}(g)]\,,
\end{eqnarray}
where we have used the relation $N_{j<i}(g)=N_g-N_{i<j}(g)$.

We now employ the relation in Eq.~(\ref{dscalc}) to derive $P_{\rm tot}(g)$.
To this end, we look upon $D_g$ as a rooted tree with the central node being the root, and use  $B_i(g)$ to denote the number of nodes in the subtree with a node at level $i$ being its root. Then, it follows that
\begin{eqnarray}\label{bi}
B_i(g) &=& \frac{2}{3}\big(N_{g-G_i(g)}-1\big)+L_{G_i(g)}(g)-i+1 \notag\\
       &=& gG_i(g)+3\cdot 2^{g-G_i(g)}-2g-i-\frac{1}{2}G_i(g)^2 \notag\\
       &&  +\frac{3}{2}G_i(g)-1.
\end{eqnarray}

For any edge $(i,j)$ connecting two nodes $i$ and $j$, we assume that node $j$ is the father of node $i$ in the rooted tree, and that node $i$ is at level $l$. Then, for the two adjacent nodes $i$ and $j$, $N_{i<j}(g) = B_l(g)$ always holds. Combining this fact and the results given by Eqs.~(\ref{dscalc}) and~(\ref{bi}), the quantity $P_{\rm tot}(g)$ can be evaluated as
\begin{eqnarray}\label{ds}
P_{\rm tot}(g) &=& \sum\limits_{i=1}^{M_g} N_i(g)B_i(g)\big[N_g-B_i(g)\big]  \notag\\
       &=& \frac{1}{8} \Big(81\cdot4^gg^2-189\cdot4^gg-189\cdot2^{2g+1}-9\cdot2^{g+1}g^3 \notag\\
       &&  +45\cdot2^{g+2}g^2+441\cdot2^{g+1}g-105\cdot2^{g+2}-52g^3 \notag\\
       &&  -120g^2+124g+816 \Big).
\end{eqnarray}

Inserting Eq.~(\ref{ds}) into Eq.~(\ref{bard}) gives
\begin{eqnarray}\label{dg}
\bar{d}_g &=& \frac{1}{8\left(81\cdot2^{2g-2}-27\cdot2^gg-45\cdot2^{g-1}+9g^2+15g+6\right)} \cdot \notag\\
          &&  \big(162\cdot4^gg^2-378\cdot4^gg-189\cdot2^{2g+2}-9\cdot2^{g+2}g^3 \notag\\
          &&  +45\cdot2^{g+3}g^2+441\cdot2^{g+2}g-105\cdot2^{g+3}-104g^3 \notag\\
          &&  -240g^2+248g+1632\big).
\end{eqnarray}
Equation~(\ref{ng}) shows that in limit of large network,
\begin{equation}\label{gn}
g \simeq \log_2N_g\,,
\end{equation}
which, together with and Eq.~(\ref{dg}), yields
\begin{equation}
\bar{d}_g \simeq \frac{162\cdot4^g g^2}{8\cdot81\cdot2^{2g-2}} \sim g^2 \sim (\ln N_g)^2
\end{equation}
for very large systems. Thus, extended dendrimers are not small-world systems, which is in contrast to compact dendrimers that display the small-world phenomenon~\cite{WaSt98} with their average path length growing logarithmically with the system size.

Actually, in addition to the average
path length, the diameter of $D_g$, denoted by $\Omega_g$, also scales with system size $N_g$ as $\Omega_g \sim (\ln N_g)^2$, which can be obtained from the following arguments. For a networked system, its diameter is defined as the maximum of the shortest distances between all pairs of nodes in the system. By construction, it is easy to prove that $\Omega_g =2 M_g = g^2-g+2$, which displays a similar scaling as that of the average
path length for large systems.

\section{Trapping in extended dendrimers with a perfect trap}\label{Trap}

After introducing the construction and structural properties of extended dendrimers, in this section we study the trapping problem in $D_g$, in order to gain a comprehensive understanding of trapping process on this dendrimer family and explore the influence of their internal structure on the trapping efficiency. We concentrate on three cases of trapping problems. In the first case, we consider trapping with a single trap positioned on the central node; for the second case, we attack trapping with the immobile trap at an arbitrary node; and in the last case, we address trapping with the trap distributed uniformly.

The trapping problem studied here is a kind of isotropic discrete-time random walks in $D_g$ with a single trap. At each time step, the walker jumps from its current location to an arbitrary nearest neighbor with identical probability. Let $T_{ij}(g)$ denote the MFPT from node $i$ to $j$. Thus, $T_{ij}(g)=0$ for $i=j$. Let $T_j(g)$ be the ATT to trap node $j$. Then, this interesting quantity is given by
\begin{equation}\label{tig}
T_j(g) = \frac{1}{N_g-1}\sum_{i} T_{i j}(g)\,.
\end{equation}
Next we will study $T_j(g)$ for the three cases of trapping problems defined in $D_g$.

\subsection{Trapping with a deep trap at the central node\label{TrapC}}

We first investigate the case that the trap is located at the central node  in $D_g$. By symmetry, all nodes at the same level have an identical TT to the trap. For the sake of brevity, we use $F_i(g)$ to represent the TT for a node at level $i$ in $D_g$. Then, $F_i(g)$ satisfies the following relations:
\begin{equation}
\label{fi}
F_i(g)=
\begin{cases}
0,                                           & i=0,  \\
\\
\frac{1}{3}[F_{i-1}(g)+1] \\
+\frac{2}{3}[F_{i+1}(g)+1], & 0<i<M_g \ {\rm and} \ i=L_{G_i(g)}(g), \\
\\
\frac{1}{2}[F_{i-1}(g)+1] \\
+\frac{1}{2}[F_{i+1}(g)+1], & 0<i<M_g \ {\rm and} \ i\not=L_{G_i(g)}(g), \\
\\
F_{M_g-1}(g)+1,                                & i=M_g.
\end{cases}
\end{equation}

For $i=0$ and $i=M_g$, Eq.~(\ref{fi}) is obvious; while for $0<i<M_g$, Eq.~(\ref{fi}) can be explained as follows. Let $p_1$ and $p_2$ represent the probabilities for a walker starting from a node at level $i$ and taking one time step to arrive at a neighboring node at level $i-1$ and level $i+1$, respectively. Since the walker performs isotropic random walks, for $i=L_{G_i(g)}(g)$, $p_1=\frac{1}{3}$ and $p_2=\frac{2}{3}$; and $p_1=p_2=\frac{1}{2}$ otherwise. For $0<i<M_g$, the first term on the right-hand side describes the process that with probability $p_1$ the walker starting from a node at level $i$ takes one step to arrive at its unique neighbor at level $i-1$, and then makes $F_{i-1}(g)$ jumps to reach the trap for the first time; the second term accounts for the fact that with probability $p_2$ the walker first makes a jump at a neighbor at level $i+1$ and then takes $F_{i+1}(g)$ steps to visit the trap for the first time.

From Eq.~(\ref{fi}), we can derive that for $0<i<M_g$,
\begin{equation}
F_i(g)-F_{i-1}(g) = 2[F_{i+1}(g)-F_i(g)] + 3
\end{equation}
and
\begin{equation}
F_i(g)-F_{i-1}(g) = F_{i+1}(g)-F_i(g) + 2
\end{equation}
hold for $i=L_{G_i(g)}(g)$ and $i\neq L_{G_i(g)}(g)$, respectively.
Let $A_i(g)=F_{i}(g)-F_{i-1}(g)$. Then,
\begin{eqnarray}
\label{ai}
A_i(g) =
\begin{cases}
2A_{i+1}(g) + 3, &  \ i=L_{G_i(g)}(g), \\
A_{i+1}(g) + 2,  &  \ i\not=L_{G_i(g)}(g),
\end{cases}
\end{eqnarray}
is true for $0<i<M_g$.
Considering $A_{M_g}(g) = F_{M_g}(g)-F_{M_g-1}(g) = 1$, Eq.~(\ref{ai}) can be solved to yield
\begin{equation}
A_i(g)=
3 \cdot 2^{-G_i(g)+g+1} -G_i(g)^2 +(2g+3)G_i(g) -2i -4g-3.
\end{equation}

Applying the obtained intermediate quantity $A_i(g)$, the TT from a node at level $i$ ($0 \leq i \leq M_g$) to the central node can be calculated by
\begin{eqnarray}\label{fig}
F_i(g) &=& F_0(g) \ + \sum\limits_{j=1}^{i} \Big [ F_j(g)-F_{j-1}(g) \Big ] \notag\\
       &=& F_0(g) \ + \sum\limits_{j=1}^{i} A_j(g)  \notag \\
       &=& \frac{1}{2}(2g+1)G_i(g)^3 -\frac{1}{4}(4g^2+8g-1)G_i(g)^2 \notag \\
       &&  +\frac{1}{4}(4g^2+4g-3)G_i(g)+(12\cdot 2^{g+2}-3\cdot2^{g+3}g)\cdot \notag\\
       &&  2^{-(G_i(g)+2)}G_i(g)+(6-3g)2^{-(G_i(g)-g-1)}     \notag \\
       &&  -G_i(g)^2i+(2g+3)G_i(g)i+3\cdot 2^{g+3}2^{-(G_i(g)+2)}i \notag \\
       &&  -i^2-4(g+1)i+3(2^{g+1}g-2^{g+2}). \notag\\
\end{eqnarray}
Then, according to Eqs.~(\ref{tig}) and~(\ref{fig}), the closed-form expression for ATT to the trap at the central node in $D_g$, denoted by $T_C(g)$, can be obtained as
\begin{eqnarray}\label{tcg}
T_C(g) &=& \frac{\sum\limits_{i=1}^{M_g} N_i(g)F_i(g)}{N_g-1}\notag \\
                     &=& \frac{1}{9 \cdot 2^{g-1}-3g-3}
                         \bigg ( 27\cdot 4^g\cdot g - 27\cdot 2^{2g+1}  \notag \\
                     &&  - 9\cdot 2^gg^3 -9\cdot 2^{g-2}\cdot g^2+45\cdot 2^{g-2}\cdot g +699\cdot 2^{g-1} \notag\\
                     &&  -14g^3-\frac{147}{2}g^2-\frac{413}{2}g-294 \bigg ). \notag\\
\end{eqnarray}
For a large system, i.e., $g\rightarrow \infty$, $T_C(g)$ has the following dominant term:
\begin{equation}
T_C(g) \simeq \frac{27\cdot 4^g\cdot g}{9 \cdot 2^{g-1}} \sim 2^g g\,,
\end{equation}
from which we can obtain the dependence relation of $T_C(g)$ on the network size $N_g$ as
\begin{equation}\label{Tcg}
T_C(g) \sim N_g\ln N_g\,,
\end{equation}
a scaling different from that for trapping in compact dendrimers, where the ATT to the center node is a linear function of the network size~\cite{WuLiZhCh12,LiZh13JCP}.

\subsection{Trapping with the trap at an arbitrary node}\label{TrapC1}

In Sec.~\ref{TrapC}, we obtained the ATT to the central node in $D_g$. Here, we use this result to further study another trapping problem with the trap located at an arbitrary node. For convenience of the following description, let $r_i$ represent a node at level $i$. Notice that all nodes at the same level are equivalent to one another, in the sense that the ATT is the same, if any of them is considered as a trap.
Let $T_{r_i}^{\rm sum}(g)$ denote the sum of the MFPT from a starting point to a target node at level $i$ $(0\leq i\leq g)$ in $D_g$, where the sum runs over all starting nodes in $D_g$. That is,
\begin{equation}\label{tris}
T_{r_i}^{\rm sum}(g) = \sum_{j \in D_g} T_{j\,r_i}(g).
\end{equation}
Then, the ATT to an arbitrary node at level $i$ in $D_g$ is
\begin{eqnarray}\label{trig}
T_{r_i}(g)=\frac{1}{N_g-1}T_{r_i}^{\rm sum}(g).
\end{eqnarray}
Thus, to find $T_{r_i}(g)$, we can alternatively evaluate the quantity $T_{r_i}^{\rm sum}(g)$.

To determine $T_{r_i}^{\rm sum}(g)$, we consider $D_g$ as a rooted tree with its root being the central node. Then, we have the following relation
\begin{eqnarray}\label{tris_calc}
&\quad&T_{r_i}^{\rm sum}(g)\notag\\
&=& \frac{2}{3}T_{r_0}^{\rm sum}(g)+\bigg[\frac{2}{3}(N_g-1)+1\bigg]T_{r_0r_i}(g)+\sum\limits_{j=1}^{i-1} T_{r_jr_i}(g) \notag\\
&&  +\sum\limits_{j=1}^{G_i(g)-1}\bigg[\frac{1}{3}T_{r_0}^{\rm sum}(g-j) + \frac{1}{3}(N_{g-j}-1)T_{r_{L_j(g)}r_i}(g) \bigg] \notag\\
&&  +\frac{2}{3}T_{r_0}^{\rm sum}(g-G_i(g))+\frac{2}{3}(N_{g-G_i(g)}-1)\cdot  \notag\\
&&  \big[ F_{L_{G_i(g)}(g)}(g)-F_i(g) \big] +\sum\limits_{j=i+1}^{L_{G_i(g)}(g)} \big[F_j(g)-F_i(g)\big], \notag\\
\end{eqnarray}
where $T_{r_j r_i}$ ($j<i$) is the MFPT from a node at level $j$ to one of its descendant node at level $i$, $F_j(g)$ (or $F_i(g)$) denotes the MFPT from a node at level $j$ (or $i$) to the central node, and $F_j(g)-F_i(g)$ ($j>i$) is thus equal to the MFPT from a node at level $j$ to its ancestor node at level $i$.

Equation~(\ref{tris_calc}) can be elaborated as follows. The first two terms describe the process that a walker starting off from a node, which and the trap node have the central node as the lowest common ancestor, should first visit the central node, and then jumps $T_{r_0 r_i}(g)$ more steps to hit the trap for the first time. Here, the lowest common ancestor for two nodes $i$ and $j$ is the node with the possible biggest level value in the rooted tree but having both $i$ and $j$ as its offspring nodes. The third and forth terms explain the sum of MFPTs to the trap, running over those starting nodes, which and the trap have the lowest common ancestor at level $j(0<j<i)$. The remaining terms account for the sum of MFPTs from all descendants of the trap to the trap itself.

We now deduce the two quantities $T_{r_0}^{\rm sum}(g)$ and $T_{r_ir_j}(g)$ with $i<j$. First, making use of Eq.~(\ref{tcg}), we can easily determine $T_{r_0}^{\rm sum}(g)$ given by
\begin{eqnarray}\label{tr0s}
T_{r_0}^{\rm sum}(g)
         &=& T_C(g)\cdot (N_g-1) \notag \\
         &=& \frac{1}{4}\big(27\cdot4^{g+1}g-27\cdot2^{2g+3}-9\cdot2^{g+2}g^3-9\cdot2^gg^2 \notag\\
         &&  +45\cdot2^gg+699\cdot2^{g+1} -56g^3-294g^2-826g \notag\\
         && -1176\big). \notag\\
\end{eqnarray}

We continue to determine $T_{r_ir_j}(g)$ $(i<j)$, which can be expressed as
\begin{eqnarray}\label{tij}
T_{r_ir_j}(g) = \sum\limits_{k=i}^{j-1} T_{r_kr_{k+1}}(g).
\end{eqnarray}
Before determining $T_{r_i r_j}(g)$, we first calculate the MFPT, $T_{r_i r_{i+1}}(g)$, from a node at level $i$($0\leq i < M_g$) to its neighboring node at level $i+1$.
For $i=0$, we have
\begin{equation}\label{t01}
T_{r_0r_1}(g)=\frac{1}{3}+\frac{2}{3}\big[1+F_1(g)+T_{r_0r_1}(g)\big]\,,
\end{equation}
which, using Eq.~(\ref{fig}), is solved to yield
\begin{equation}\label{t01_val}
T_{r_0r_1}(g)=3\cdot 2^{g+1}-4g-3.
\end{equation}
For $0<i<M_g$, the following relations can be established:
\begin{small}
\begin{eqnarray}\label{tii+1}
&&T_{r_ir_{i+1}}(g)\notag\\
&=&
\begin{cases}
\frac{1}{3}+\frac{1}{3}\big[1+T_{r_{i-1}r_i}(g)+T_{r_ir_{i+1}}(g)\big] \\
+\frac{1}{3}\big[1+F_1(g-i)+T_{r_ir_{i+1}}(g)\big], & i=L_{G_i(g)}(g), \\
\\
\frac{1}{2}+\frac{1}{2}\big[1+T_{r_{i-1}r_i}(g)+T_{r_ir_{i+1}}(g)\big], & i\not=L_{G_i(g)}(g).
\end{cases}\notag\\
\end{eqnarray}
\end{small}
Considering the initial condition in Eq.~(\ref{t01_val}), Eq.~(\ref{tii+1}) is solved to give
\begin{small}
\begin{eqnarray}\label{tii+1_val}
T_{r_ir_{i+1}}(g)=
\begin{cases}
9\cdot2^g-4g -3\cdot2^{g-G_i(g)} \\
-2G_i(g) -3, & i=L_{G_i(g)}(g),\\
\\
9\cdot2^g-3 -2g    \\
-3\cdot2^{g-G_i(g)+1}-2gG_i(g) \\
+G_i(g)^2-3G_i(g)+2i-1, & i\not=L_{G_i(g)}(g).
\end{cases}
\end{eqnarray}
\\\\\\
\end{small}
Substituting Eq.~(\ref{tii+1_val}) into Eq.~(\ref{tij}) leads to
\begin{eqnarray}\label{tijg_val}
T_{r_ir_j}(g) &=& \sum\limits_{k=i}^{j-1} T_{r_kr_{k+1}}(g) \notag\\
           &=& \left(3\cdot2^{g-G_i(g)}-g^2-2g+\frac{1}{4}\right)G_i(g)^2+ \bigg(g^2 \notag\\
           &&  -3g\cdot2^{g-G_i(g)+1}+9\cdot2^{g-G_i(g)}+g-\frac{1}{2}\bigg)G_i(g) \notag\\
           &&  +\left(g^2-3\cdot2^{g-G_j(g)}+2g-\frac{1}{4}\right)G_j(g)^2+\bigg(-g^2 \notag\\
           &&  +3g\cdot2^{g-G_j(g)+1}-9\cdot2^{g-G_j(g)}-g+\frac{1}{2}\bigg)G_j(g) \notag\\
           &&  +\big[(2g+3)G_i(g)+3\cdot2^{g-G_i(g)+1}-9\cdot2^g+2g \notag\\
           &&  -G_i(g)^2+2\big]i+\left(g+\frac{1}{2}\right)G_i(g)^3+\big[-(2g+3)G_j(g) \notag\\
           &&  -3\cdot2^{g-G_j(g)+1}+9\cdot2^g-2g+G_j(g)^2-2\big]j \notag\\
           &&  +3\cdot2^{g-G_i(g)+2}-3g\cdot2^{g-G_i(g)+1}-\left(g+\frac{1}{2}\right)G_j(g)^3 \notag\\
           &&  -3\cdot2^{g-G_j(g)+2}+3g\cdot2^{g-G_j(g)+1}-i^2+j^2 \notag\\
           &&  -\frac{1}{4}G_i(g)^4+\frac{1}{4}G_j(g)^4. \notag\\
\end{eqnarray}

Finally, plugging Eqs.~(\ref{fig}),~(\ref{tr0s}), and~(\ref{tijg_val}) into Eq.~(\ref{tris_calc}), then inserting the obtained result into Eq.~(\ref{trig}), we analytically obtain a rigorous expression for the ATT to any trap node at level $i$ of the extended dendrimer $D_g$:
\begin{small}
\begin{widetext}
\begin{eqnarray}\label{trig2}
T_{r_i}(g)
&=& \frac{T_{r_i}^{\rm sum}(g)}{N_g-1} \notag\\
&=& \frac{1}{9\cdot2^{g-1}-3g-3} \cdot \notag\\
&& \Big[ \big(-9\cdot2^g-6G_i(g)^2+6G_i(g)-14\big)g^3+\big(9\cdot2^gG_i(g)^2-9\cdot2^gG_i(g)-9G_i(g)2^{g-G_i(g)+2}-9\cdot2^{g-G_i(g)+2}+135\cdot2^{g-2} \notag\\
&&  +12iG_i(g)-6i+6G_i(g)^3-17G_i(g)^2+11G_i(g)-\frac{147}{2}\big)g^2 + \big(-9\cdot2^{g+1}iG_i(g)+9i2^{g-G_i(g)+2}-9\cdot2^{g+1}i \notag\\
&&  -9\cdot2^gG_i(g)^3+9\cdot2^{g+1}G_i(g)^2+9G_i(g)^22^{g-G_i(g)+1}-9\cdot2^gG_i(g)+3G_i(g)2^{g-G_i(g)+3}+27G_i(g)2^{2g-G_i(g)+1} \notag\\
&&  +21\cdot2^{g-G_i(g)+1}+27\cdot2^{2g-G_i(g)+1}-123\cdot2^{g-2}-27\cdot2^{2g}-6i^2-6iG_i(g)^2+\notag\\
&&
28iG_i(g)-11i-\frac{3}{2}G_i(g)^4+8G_i(g)^3-\frac{17}{2}G_i(g)^2-2G_i(g)-\frac{413}{2}\big)g
+9\cdot2^gi^2 + 9\cdot2^giG_i(g)^2-27\cdot2^giG_i(g)
\notag\\
&&
+15i2^{g-G_i(g)+1}-27i2^{2g-G_i(g)+1} \notag\\
&&  -27\cdot2^{g-1}i+81\cdot2^{2g-1}i+9\cdot2^{g-2}G_i(g)^4-9\cdot2^{g-1}G_i(g)^3-9\cdot2^{g-2}G_i(g)^2+15G_i(g)^22^{g-G_i(g)}-27G_i(g)^22^{2g-G_i(g)} \notag\\
&&  +9\cdot2^{g-1}G_i(g)+45G_i(g)2^{g-G_i(g)}-81G_i(g)2^{2g-G_i(g)}+15\cdot2^{g-G_i(g)+2}-27\cdot2^{2g-G_i(g)+2}+579\cdot2^{g-1} \notag\\
&&   +27\cdot2^{2g+1}-5i^2-5iG_i(g)^2+15iG_i(g)-5i-\frac{5}{4}G_i(g)^4+\frac{5}{2}G_i(g)^3+\frac{5}{4}G_i(g)^2-\frac{5}{2}G_i(g)-294 \Big]. \notag\\
\end{eqnarray}
\end{widetext}
\end{small}

In this way, we have obtained an explicit formula for the ATT to an arbitrary node in $D_g$, as given by Eq.~(\ref{trig2}). For the particular case $i=0$ corresponding to the trapping with the trap placed at the central node, Eq.~(\ref{trig2}) reduces to the expression in Eq.~(\ref{tcg}). For another limiting case $i=M_g$ that the trap is located at a peripheral node, we use $T_{P}(g)$ to denote the ATT. For this case, plugging $i=M_g=(g^2-g+2)/2$ into Eq.~(\ref{trig2}) yields
\begin{eqnarray}
T_{P}(g) &=& \frac{1}{9\cdot2^{g-1}-3g-3}\Big(81\cdot2^{2g-2}g^2-189\cdot2^{2g-2}g+ \notag\\
       &&  189\cdot2^{2g-1} -27\cdot2^gg^3+81\cdot2^{g-1}g^2-147\cdot2^{g-1}g \notag\\
       &&  +123\cdot2^g+3g^4-\frac{29}{2}g^3-58g^2-\frac{191}{2}g-214\Big). \notag\\
\end{eqnarray}
For very large systems ($g \rightarrow \infty$), we have
\begin{equation}\label{Tpg}
T_{P}(g) \simeq \frac{81\cdot2^{2g-2}g^2}{9\cdot2^{g-1}-3g-3} \sim 2^gg^2 \sim N_g\big(\ln N_g\big)^2\,,
\end{equation}
which is also different from the $N_g\ln N_g$ scaling of ATT for trapping in compact dendrimers with a peripheral node being the trap~\cite{WuLiZhCh12,LiZh13JCP}.

\begin{figure}
\begin{center}
\includegraphics[width=1.10\linewidth,trim=15 30 0 35]{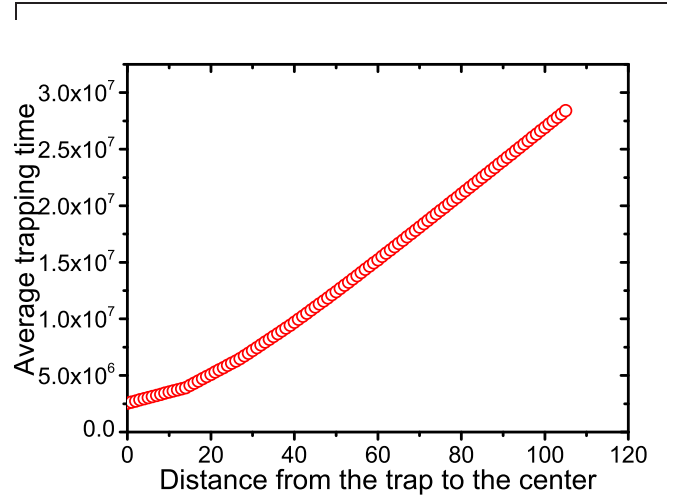}
\end{center}
\caption[kurzform]{Average trapping time for the trap node at different levels in  $D_{15}$. }\label{ATT}
\end{figure}

For general case of $0<i<M_g$, the expression for ATT provided in Eq.~(\ref{trig2}) is rather
lengthy and is hard to observe the the dependence relation of ATT on the system size $N_g$. In order to show how the ATT scales with $N_g$, In Fig.~\ref{ATT}, we report the results, given in Eq.~(\ref{trig2}), for trapping in $D_{15}$ with the trap positioned at nodes having different distances towards to central node. Figure~\ref{ATT} shows that the ATT depends on the level of trap node, which increases with the distance between the central node and the trap, i.e., $T_{r_i}(g)\sim (i+1)N_g\ln N_g$. Note that for trapping in compact dendrimers~\cite{LiZh13JCP}, the ATT also grows with the level of the trap but displays a different behavior from that corresponding to extended dendrimers.

\subsection{Trapping with the trap uniformly distributed}

We are now in position to study trapping in extended dendrimer $D_g$ with the trap uniformly distributed throughout all nodes. In this case, the ATT, denoted by $\langle T\rangle_g$ is defined as the average of MFPTs over all pairs of nodes in $D_g$ :
\begin{eqnarray}\label{E00}
\langle T\rangle_g=\frac{1}{N_g(N_g-1)}\sum_{i=1}^{N_g}\sum_{j=1}^{N_g} T_{ij}(g)\,.
\end{eqnarray}
For convenience, let $T_{\rm tot}(g)$ denote the summation term on the rhs of Eq.~(\ref{E00}), that is,
\begin{equation}\label{E01}
T_{\rm tot}(g)=\sum_{i=1}^{N_g}\sum_{j=1}^{N_g} T_{ij}(g)\,.
\end{equation}
Then,
\begin{equation}\label{E000}
\langle T\rangle_g=\frac{T_{\rm tot}(g)}{N_g(N_g-1)}\,.
\end{equation}
By definition, $\langle T\rangle_g$ involves a double average: The first one is the average of trapping times  running over all starting nodes to a given trap node, the second one is the average of the first one with the trap having a uniform distribution. In what follows, we will apply two disparate approaches to analytically determine $T_{\rm tot}(g)$ and $\langle T\rangle_g$, with the results obtained by both methods being consistent with each other.

We first determine $T_{\rm tot}(g)$ and $\langle T\rangle_g$ by using the intermediary results obtained in Section~\ref{TrapC1}. For $T_{\rm tot}(g)$, it satisfies
\begin{equation}\label{ttotg}
T_{\rm tot}(g) = \sum\limits_{i=0}^{M_g} N_i(g)\big[T_{r_i}(g)(N_g-1)\big]\,.
\end{equation}
Inserting Eq.~(\ref{ttotg}) into Eq.~(\ref{E000}) and using Eq.~(\ref{trig2}), we obtain
\begin{small}
\begin{eqnarray}\label{tg_res}
\langle T \rangle_g&=& \frac{\sum\limits_{i=0}^{M_g} N_i(g)\big[T_{r_i}(g)(N_g-1)\big]}{N_g(N_g-1)}\notag\\
 &=&  \frac{1}{81\cdot2^{2g-2}-27\cdot2^gg-45\cdot2^{g-1}+9g^2+15g+6} \cdot \notag\\
         \notag\\
    &&   \Big(729\cdot8^{g-1}g^2-1701\cdot8^{g-1}g-1701\cdot2^{3g-2}-81\cdot4^gg^3 \notag\\
    &&   -1323\cdot2^{g-1}g^2+567\cdot2^{2g-1}g^2+2835\cdot2^{2g-1}g \notag\\
    &&   -189\cdot4^g+27\cdot2^{g-1}g^4-45\cdot2^{g+2}g^3 \notag\\
    &&   -540\cdot2^{g-1}g^2-207\cdot2^gg+1233\cdot2^g+39 g^4 \notag\\
    &&   +129g^3-3g^2-705g-612\Big)\,. \notag\\
\end{eqnarray}
\end{small}

Besides the above method, we can also compute $T_{\rm tot}(g)$ and $\langle T\rangle_g$ by using the connection between effective resistance and MFPTs for random walks~\cite{Te91}. For this purpose, we view $D_g$ as an electrical network~\cite{DoSn84} by considering every edge in $D_g$ to be a unit resistor~\cite{KlRa93}. Let $R_{ij}(g)$ denote the effective resistance between a pair of nodes $i$ and $j$ in the electrical network associated with $D_g$. Then, the following relation holds
\begin{equation}\label{H00}
T_{ij}(g)+T_{ji}(g)=2E_g\, R_{ij}(g)\,.
\end{equation}
Using this link governing MFPTs and effective resistance, Eq.~(\ref{E01}) can be rewritten as
\begin{equation}\label{H01}
T_{\rm tot}(g)=E_g\sum_{i=1}^{N_g}\sum_{j=1}^{N_g}R_{ij}(g)\,.
\end{equation}
Since extended dendrimers have a treelike structure, the effective resistance $R_{ij}(g)$ is exactly the usual shortest-path distance $d_{ij}(g)$ between nodes $i$ and $j$ in $D_g$. Then, making use of Eqs.~(\ref{bard}) and~(\ref{dsdf}), we have
\begin{equation}\label{H02}
T_{\rm tot}(g)=E_g\sum_{i=1}^{N_g}\sum_{j=1}^{N_g}d_{ij}(g)=E_g\,P_{\rm tot}(g)
\end{equation}
and
\begin{equation}\label{H03}
\langle T \rangle_g = \frac{T_{\rm tot}(g)}{N_g(N_g-1)} = \frac{E_g P_{\rm tot}(g)}{N_g(N_g-1)} = E_g \cdot \bar{d}_g\,.
\end{equation}
Inserting Eq.~(\ref{dg}) into Eq.~(\ref{H03}), we recover Eq.~(\ref{tg_res}).

When the system size is large enough, the ATT $\langle T \rangle_g$ in Eq.~(\ref{tg_res}) can be approximated as
\begin{equation}\label{H04}
\langle T \rangle_g \simeq \frac{729\cdot8^{g-1}g^2}{81\cdot2^{2g-2}} \sim 2^g g^2\sim N_g\big(\ln N_g\big)^2\,,
\end{equation}
which differs greatly from the leading asymptotical scaling for ATT in compact dendrimers with a trap distributed uniformly among all nodes~\cite{LiZh13JCP}.

\subsection{Result comparison and analysis}

From the above obtained results, provided by Eq.~(\ref{Tcg}), Eq.~(\ref{Tpg}), and Fig.~\ref{ATT}, it can be seen that the ATT for trapping in extended dendrimers exhibits rich scalings when the trap is placed at different positions. Figure~\ref{ATT} and Eq.~(\ref{trig2}) show that the trap's location has a significant effect on the efficiency of trapping in extended dendrimers measured by ATT: The trapping efficiency decreases when the distance from the trap to the central node increases. For the case that central node is the trap, the trapping process is the most efficient, its efficiency scales with system size $N_g$ as $N_g \ln N_g$; while for the case that a peripheral node is the trap, the trapping process is the least efficient, the efficiency varies with network size $N_g$ as $N_g(\ln N_g)^2$.

The different behaviors for $T_{r_i}(g)$ provided in Eq.~(\ref{trig2}) lie in the particular structure of the extended dendrimers. Figure~\ref{Cayley} shows that the extended dendrimer under consideration consists of $3$ branches (regions), each being a subtree with a node at level $1$ as its root. For $i=0$ corresponding to the case that the central node is the trap, the walker, regardless of its starting position, will visit at most one branch before being absorbed by the trap. For $i>0$ corresponding the case when the trap is placed on a node with distance $i$ to the center, the walker, starting from a large group of nodes, must first visit the central node and then proceeds from the center along the path $r_{0}- r_{1}-\cdots -r_{i-1}-r_{i}$ until it is trapped. Equation~(\ref{tii+1_val}) shows that the MFPT $T_{r_{i-1}r_i}$ from a node at level $i-1$ to its direct neighbor at level $i$ grows with $i$. Therefore, the scaling of $T_{r_i}(g)$ grows with increasing $i$. In particular, when the trap is moved away from the central node to a peripheral node, the ATT changes from $N \ln(N)$ to $N \ln(N)^2$.

On the other hand, when the trap is uniformly distributed over $D_g$, the dominant scaling of ATT $\langle T \rangle_g$ scales with the network size $N_g$ as $N_g(\ln N_g)^2$, which is the same as that of $T_{P}(g)$ for trapping in $D_g$ with a trap fixed at a peripheral node. Thus, the $N_g \ln N_g$ behavior of the ATT to the central node is not representative of extended dendrimers, while the scaling of ATT to a peripheral node is a representative property for trapping in extended dendrimers. The equivalence of the dominant scalings of ATTs for $T_{P}(g)$ and $\langle T \rangle_g$ can be easily understood. From Eqs.~(\ref{ng}) and (\ref {Np}), it is evident that in $D_{g}$ the fraction of peripheral nodes is about $\frac{1}{3}$. Furthermore, the ATT is the highest for all trapping problems in $D_g$ with a single immobile trap.
Therefore, the $T_{P}(g)$ alone determines the leading  behavior of $\langle T \rangle_g$.

The aforementioned results also indicate that for the three cases of trapping problems under consideration,  the trapping processes are less efficient in extended dendrimers than in compact dendrimers. 
Actually, trapping in extended dendrimers also displays different phenomena from other complex systems. For example,
for trapping in Vicsek fractals~\cite{LiZh13JCP,Pe14} as a model of regular hyperbranched polymers~\cite{BlJuKoFe03,BlFeJuKo04,FuDoBl13}, the efficiency is identical, despite of the location of the trap. Furthermore, trapping process in extended dendrimers is more efficient than in Vicsek fractals, implying that extended dendrimers have a desirable structure favorable to diffusion, as opposed to Vicsek fractals.
In addition to compact dendrimers and Vicsek fractals, trapping has also been studied in various other trees, such as $T-$shape treelike fractals and their extensions~\cite{KaRe89,Ag08,HaRo08,LiWuZh10,ZhWuCh11,JuWuZh13,PeXu14}, fractal scale-free trees~\cite{ZhLiMa11}, small-world uniform recursive trees~\cite{ZhQiZhGoGu10,ZhLiLiCh11,LiuZh13}, non-fractal scale-free trees~\cite{ZhLiMa11}. Previously reported results show that trapping behaviors in the first two network families are similar to those of Vicsek fractals, and that  trapping behaviors in the last two network families resemble those of compact dendrimers.
Thus far, the dependence relation $N_g(\ln N_g)^2$ of $T_{P}(g)$ and $\langle T \rangle_g$ on system size $N_g$ has not been observed in other systems, implying that the structure of extended dendrimers is unique.

Before closing this section, we should stress that although we only focus on trapping in a particular extended dendrimer, where any branching site has three nearest neighbors, for trapping in other
extended dendrimers with the number of nearest neighbors of each branching site greater than 3, the qualitative scalings are similar to those found for the specific extended dendrimer under consideration.

\section{Conclusions}

We have performed an extensive study of the trapping problem on extended dendrimers with varying lengths of the molecular branches. We first addressed a particular case with the trap fixed at the central node; we then considered the case that the trap is positioned at an arbitrary node; and we finally studied the case with the trap uniformly distributed over all nodes in the systems. For these three cases of trapping problems, we studied the ATT as an indicator of the trapping efficiency, and obtained  closed-form formulas for the ATTs, as well as their leading terms. We found that the location of trap has an essential influence on the trapping efficiency, with the ATT increasing with the shortest distance between the central node and the trap. We showed that the behavior of trapping in extended dendrimers is unusual, since the leading scalings of the ATTs are unique, which are not observed in compact dendrimers and even other complex networked systems.

\begin{acknowledgments}
This work was supported by the National Natural Science Foundation of China under Grant No. 11275049.
\end{acknowledgments}

\nocite{*}

\end{document}